# ON THE FIELD DEPENDENCE OF THE INTERFACE ENERGY IN AF/FM BILAYERS

J.R.L. de Almeida , J.R.Steiner, S.M.Rezende

*Departamento de Física, Universidade Federal de Pernambuco, 50670-901, Recife-PE, Brazil*

**Abstract**

In the investigations of antiferromagnetic (AF)/ ferromagnetic (FM) bilayer samples, often distinct experimental techniques yield different values for the measured exchange anisotropy field ($H_E$). We propose that the observed discrepancy may be accounted in part by the dependence of the unidirectional anisotropy with the value of the external applied field ($h$). Using a simple microscopic model for representing the AF/FM interface, which incorporates the effect of interface roughness, we show that the interface energy between the AF and FM layer indeed varies with h, as recently observed in anisotropic magnetoresistance measurements, lending support to our proposal.

Exchange bias or unidirectional anisotropy, which is characterized among other effects by the offset in the field of the magnetic hysteresis loop from zero, has become in recent years an important issue both technologically and in condensed matter physics. A thorough understanding of all variables at play still is a debatable topic in discussions about the involved phenomena (see the reviews in [1-4]). It seems that due to the diversity of systems exhibiting the phenomenon there may be more than just one mechanism to generate it. Theoretical models have considered both compensated and uncompensated interfaces, single-crystal and polycrystalline systems, spin-flop coupling, interface roughness, and magnetic domains in the antiferromagnetic layer [5]. Up to now there is not a definitive model to account for all the richness of effects observed. In this work we employ a simple Ising model [6] which display many of the hallmarks of systems exhibiting uniaxial anisotropy to study the external field dependence of the interface energy. In figure 1a we illustrate the usual model for an uncompensated AF interface and interlayer interactions, and in figure 1b for a compensated AF interface with ferromagnetic interlayer interaction, both including the effect of disorder or roughness at the interface. It is clear for the case in figure 1a that the directional symmetry is broken as long as the AF structure is not inverted by the external field. In the case of figure 1b the broken directional symmetry may look a little more subtle because it depends crucially on the interface disorder: broken directional symmetry looks like the one illustrated in figure 1c where the heights may be seen as local magnetizations of the AF or FM.

It is well known that in the investigation of samples of antiferromagnetic(AF)/ferromagnetic(FM) bilayers, often distinct experimental techniques yield different values for the measured exchange anisotropy field ($H_E$). This intriguing fact has been interpreted as arising from the distinct natures of the experimental techniques, some probe reversible properties of the system while others probe irreversible properties. Here we propose another reason for the observed discrepancy, namely the dependence of the interface energy between the AF and FM layers with the value of the external applied field ($h$), and the fact that each experimental technique employs a different field range. We use further simplification of the model represented in figure 1 but which still retains its essential features such as incorporating the effect of roughness at the AF/FM interface [6].. This model has been used earlier to study the thermal-history-dependent properties observed in exchange-coupled AF/FM bilayers. Consider two atomic monolayers with magnetic moments over congruent square lattices, one layer with ferromagnetically coupled moments and the other with two perfectly compensated antiferromagnetic sublattices. The moments from different layers are coupled by an interlayer exchange interaction, which can be FM or AF. The interface roughness is accounted for by randomly substituting a fraction of the atoms in the FM layer by atoms from the AF layer. The system Hamiltonian is taken as

$$H = H_{AF} + H_{FM} + H_c \tag{1}$$

where $H_{AF}, H_{FM}$ and $H_c$ are, respectively, the interaction energies in the antiferromagnetic layer (AFL), in the ferromagnetic layer (FML) and the coupling between the FML and AFL atoms. Thus we have

$$H_{AF} = -\sum_{(ij)} J_{ij}^{(1)} \sigma_i^{(1)} \sigma_j^{(1)} - D_1 \sum_i \sigma_i^{(1)2} - h\sum_i \sigma_i^{(1)} \tag{2}$$

where $\sigma_i^{(1)} = S, S-1,...,-S$ means the spins on the AFL at site i interacting through nearest neighbor (NN) exchange interaction $J_{ij}^{(1)} = J_1 < 0$, $(ij)$ meaning sum over all NN pairs, $D_1$ is a local uniaxial anisotropy in the AFL, and $h$ is the external field. Let the local randomly distributed variables $\eta_i = 1,0$ specify the presence (=1) or absence (=0) of a FML atom at site i which, in the latter case, is assumed substituted by a AFML atom. Hence, for the FM layer,

$$H_{FM} = -\sum_{(ij)} J_{ij}^{(2)} S_i S_j - \sum_i \left[ D_2 \eta_i S_i^2 + D_1(1-\eta_i)(\sigma_i^{(2)})^2 \right] - h\sum_i \left[ (1-\eta_i)S_i + \eta_i \sigma_i^{(2)} \right]$$

(3)

where $S_i = S, S-1,...,-S$ means the spins on the FML at site i interacting through nearest neighbors (NN) exchange interaction $J_{ij}^{(2)}$, $D_2$ is a local uniaxial anisotropy in the FML, and $\sigma_i^{(2)}$ denotes the moment of an AFL atom in FM layer. Due to the random substitution of FML atoms by AFL ones, $J_{ij}^{(2)}$ assumes the form,

$$J_{ij}^{(2)} = J_2 \eta_i \eta_j - J_1(1-\eta_i)(1-\eta_j) + J_c\left[(1-\eta_i)\eta_j + \eta_i(1-\eta_j)\right].$$ (4)

The interlayer exchange interaction is,

$$H_c = -\sum_i \left[ J_c \eta_i S_i \sigma_i^{(1)} - J_1(1-\eta_i)\sigma_i^{(1)}\sigma_i^{(2)} \right]$$ (5)

where the sum is over all sites at the interface and $J_c$ represents the coupling between FML and AFL atoms. All energies shall be measured in units of $J_1$. The many-body problem posed by the model expressed by (1) - (5) is far from trivial. Like as in random field magnets and spin glasses ( behaviours also observed in some exchange biased systems [7] with even a typical spin glass Almeida-Thouless line [8] being detected [9] ), the presence of randomness results in a complex phase space with strong metastability effects always present. The last terms in equations (3) and (4) act like an effective random field in the system and explicitly breaks time-reversal symmetry in the ferromagnetic sub-system, i.e., as long as the AF structure is not altered by the external field the FM structure is not invariant under change of magnetization direction giving origin to the unidirectional anisotropy as argued in [10] in terms of random fields acting in the interface. Following the mean-field approach of Soukoulis et al [11] in their studies of random magnets, the local thermally averaged magnetization at temperature $T$ is given by

$$M_i^{(\mu)} = \sum_{\sigma=1}^{S} \sinh(\phi_i^{(\mu)}(\sigma))\exp(D\sigma^2) \bigg/ \left[ \sum_{\sigma=1}^{S} \cosh(\phi_i^{(\mu)}(\sigma))\exp(D\sigma^2) + 0.5 \right] \quad (6a)$$

where

$$\phi_i^{(\mu)}(\sigma) = \beta\sigma \sum_{\mu} \sum_{(ij)} (J_{ij} M_j^{(\mu)} + h) \quad (6b)$$

is the local field in the mean field approximation, $\mu=1,2$ specify, respectively, FML or AFML atoms, the sum $(ij)$ is over NN, $J_{ij}$ is defined in equation (4) and $\beta = 1/T$ is in units of Boltzmann constant. Following the same procedure as in [6,11], equations (6a) and

(6b) are solved numerically by an iterative method, yielding the local and macroscopic magnetizations. As should be expected from the set of nonlinear equations (6), there are many possible local arrangements for the spins and thus effects of irreversibility and metastability sets in. As in most experimentally studied systems we have chosen $J_1, J_2, J_c, D_1$ and $D_2$, such that $T_N < T_C$. Equations (6) are iterated for a given (cooling) field $h$ in an initial $T_i$ in the range $T_N < T_i < T_C$ and the temperature is decreased in steps $\Delta T$, to a measuring final temperature $T$. Then the field can be varied, with $T$ kept constant, for obtaining the hysteresis loop. Considering an AF/FM compensated bilayer with parameters $J_1 = -1.00$, $J_2 = 1.20$, $J_c = 0.80$ $D_1 = 1.00, D_2 = 0.00$, roughness parameter $p =$ 0.15 (= 1 – mean($\eta_i$)) ; in figures 2a and 2b it is shown the resulting five hysteresis cycles at T=6.0 cooling from an initial temperature T=9.0, spin quantum number S=2, and cooling fields 0.40 and 0.25, respectively. The blocking temperature is $T_B \approx 7.5$ in this case. As can be seen in the figures both training and memory effects are present ( throughout we have considered for the AF/FM system two square monolayers of size 2 x 100 x 100). The interlayer exchange energy is given by equation 3 and in figures 3a ,3b (S=1 case) and 4a,4b (S=2 case) we obtain its field dependence for several values of the interlayer exchange coupling. As may have been expected this quantity is field dependent, the results in figures 3a and 4a agrees qualitatively with the experimentally observed results [12]. However, new theoretical results are shown in Fig. 3 (b) and Fig. 4 (b), where we have the prediction of an inversion of behavior of the interlayer energy when the interlayer coupling is varied. Qualitatively, the interface energy depends on the relative orientation of the magnetic moments at the AF and FM and the applied external field. As the latter changes this energy may increase or decrease depending on interface exchange coupling and field intensity. So distinct experimental techniques should yield different values for the measured exchange anisotropy field if not all parameters applied to the physical system are identical.

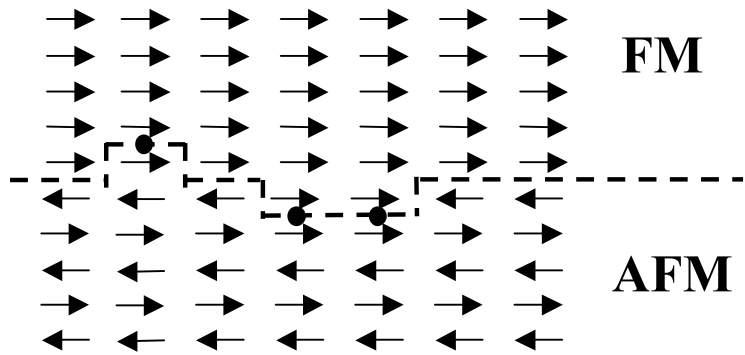

FIG. 1a: Rough interface with frustrated interactions marked by full dots. The dashed line marks the boundary between the FM and AFM

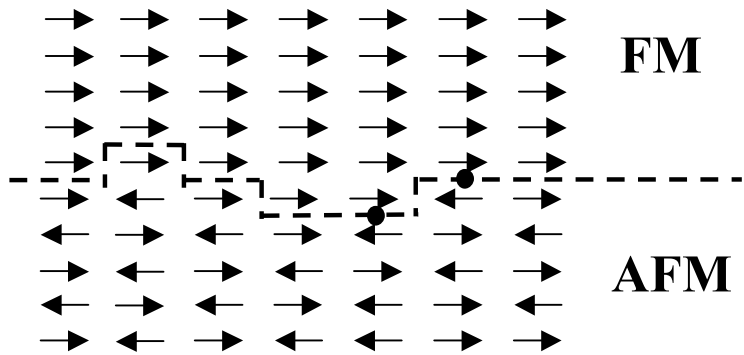

FIG. 1b: Rough interface with frustrated interactions marked by full dots. The dashed line marks the boundary between the FM and AFM

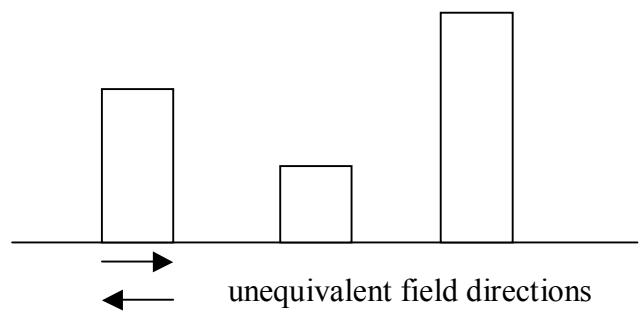

FIG. 1c: artistic view of broken unidirectional symmetry (see text)

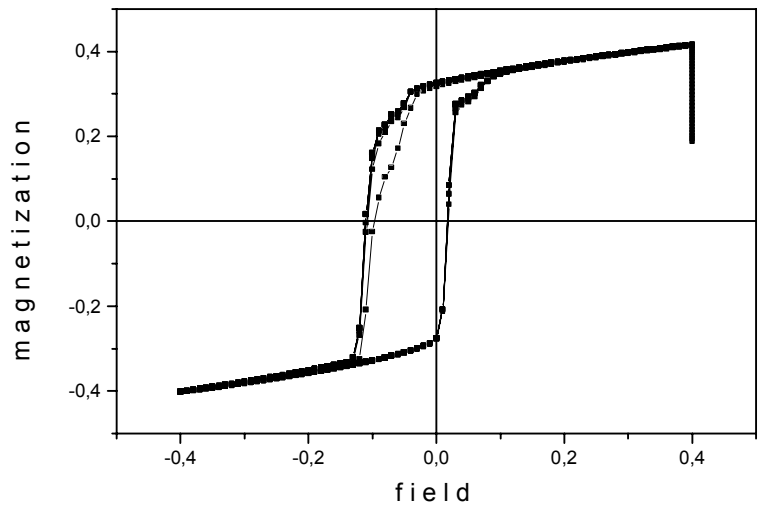

Figure 2a: hysteresis cycles , initial cooling field set at 0.40 (see text)

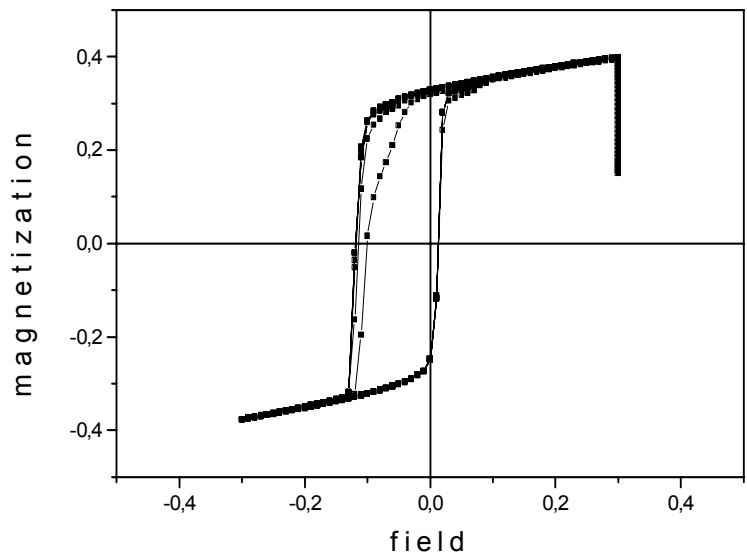

Figure 2b: hysteresis cycles , initial cooling field set at 0.25 (see text)

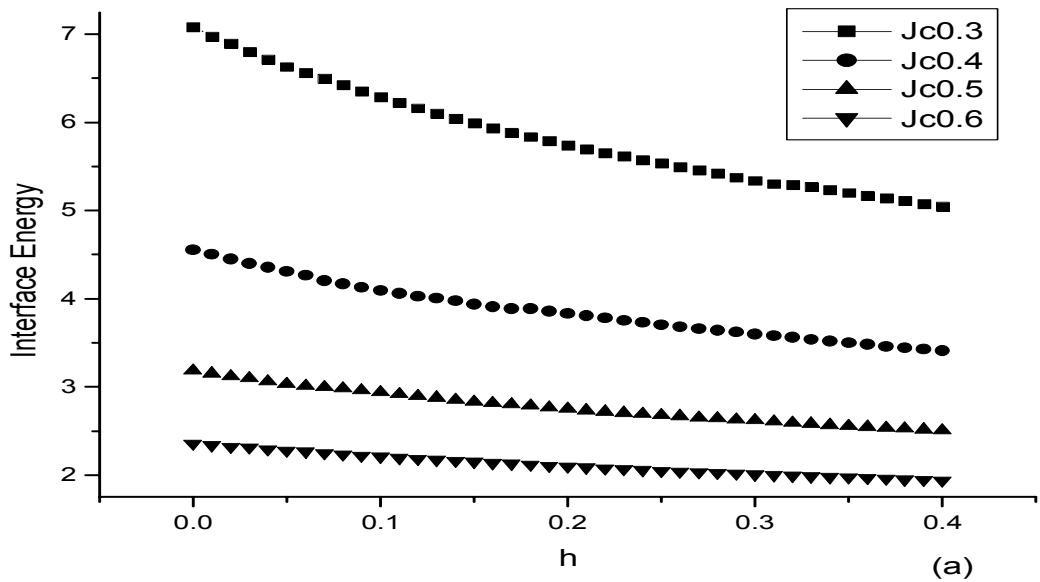

Figure 3a: Interface energy between the AF and FM layers as function of the applied field for S=1 systems

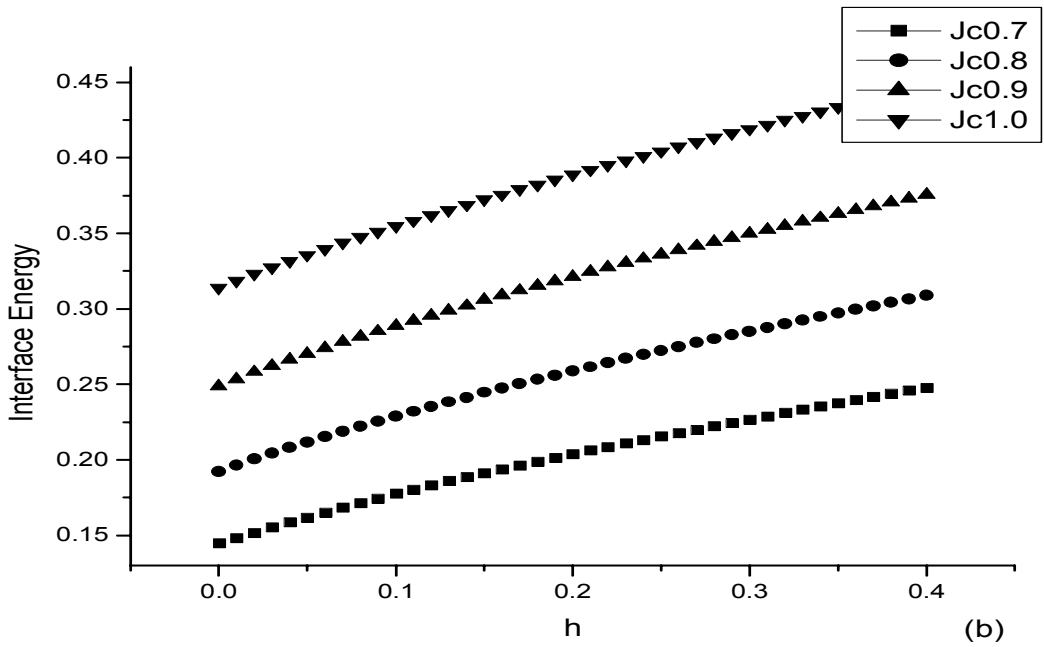

Figure 3b: Interface energy between the AF and FM layers as function of the applied field for S=1 systems Fig. 2: Interface energy between the AF and FM layers as function of the applied field for S=2 systems

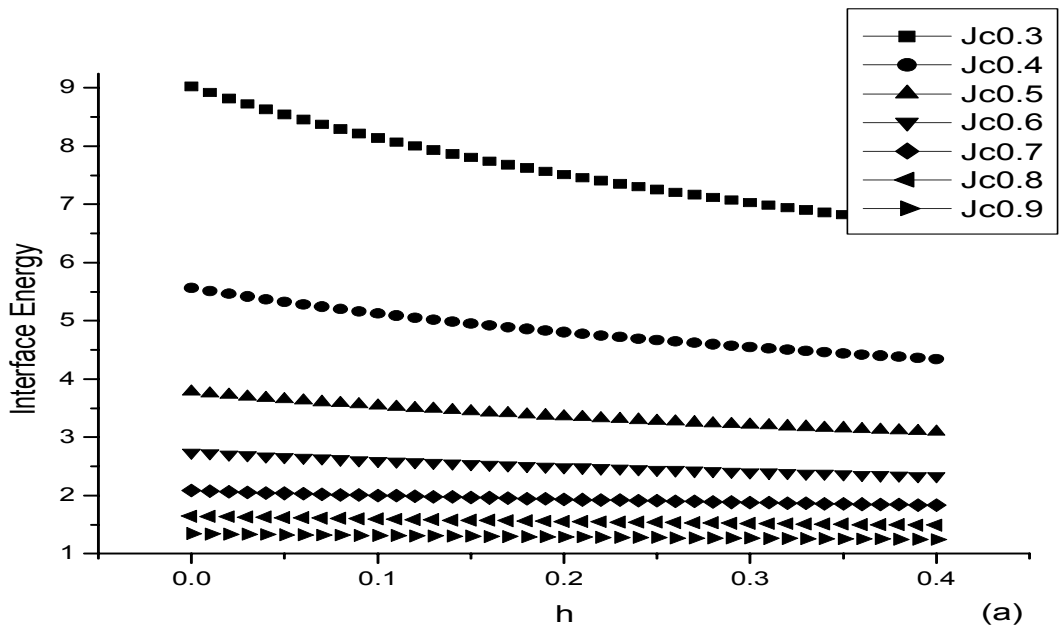

Figure 4a : Interface energy between the AF and FM layers as function of the applied field for S=2 systems

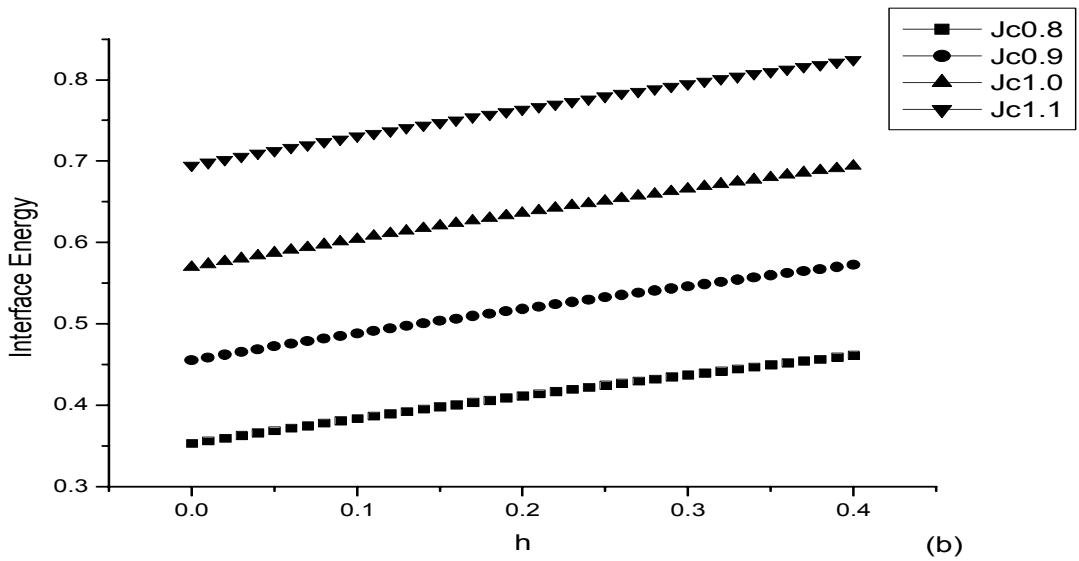

Figure 4b : Interface energy between the AF and FM layers as function of the applied field for S=2 systems